\documentclass[twocolumn,superscriptaddress,aps]{revtex4}%
\usepackage{epsfig,dsfont,amssymb,amsmath,amsthm,amsfonts,amsbsy,mathrsfs}
\usepackage{graphicx}
\usepackage{amsmath}
\usepackage{amssymb}
\usepackage{amsfonts}%
\begin{document}
\title{Efficient graph state generation and operation error detection with
controlled-path gate}
\author{Qing Lin}
\email{qlin@mail.ustc.edu.cn}
\affiliation{College of Information Science and Engineering, Huaqiao University (Xiamen),
Xiamen 361021, China}
\affiliation{Institute for Quantum Information Science, University of Calgary, Alberta T2N
1N4, Canada}
\author{Bing He}
\email{heb@ucalgary.ca}
\affiliation{Institute for Quantum Information Science, University of Calgary, Alberta T2N
1N4, Canada}


\begin{abstract}
How to generate photonic graph states efficiently is the main problem in
optical measurement-based quantum computation. We propose a deterministic
approach based on an element two-qubit gate called the controlled-path gate
(Phy. Rev. A 80, 042310 and Phys. Rev. A 80, 042311) for the purpose. For the
generation of the wheel-shape graph states, the procedure is especially
simplified compared with the other methods. Moreover, we consider the
possibility of finding the operation errors occurring in generating the graph
states of large numbers of qubits; the use of controlled-path gate allows to
detect with a certain probability if there is a missing of entangling
operation for the input qubits.

\end{abstract}
\maketitle

\section{Introduction}

The measurement-based quantum computation (MBQC) or one-way quantum
computation has attracted more and more attentions since the idea was
introduced \cite{cluster1, cluster2, cluster3}. Its main difference from the
traditional circuit-based quantum computation is that only single-qubit
measurements are implemented on the apriori prepared graph states. The
difficulty of constructing the multi-qubit gates in circuit-based approaches
is therefore traded off to that of preparing the specified graph states in a
computation task. It is conceivable that, for the off-line preparation of
graph states, multi-qubit operations should be necessary to entangle the input
qubits. Numerous theoretical (see e.g. \cite{Nielsen1, Browne, Gilbert,
dc-c1}) and experimental studies (see e.g. \cite{experiments1, experiments2,
k, pan1, y, pan2}) have been devoted to creating the graph states of photons
for optical MBQC.

Most of the previous schemes for preparing graph states work in the pattern of
entangling one qubit after another. To facilitate the efficient generation of
the photonic graph states, there have been the proposals of constructing such
states from various building blocks \cite{qlin6} and bundling the entangling
operations to a whole operation \cite{qlin5}. Here we provide a different
method to bundle all entangling operations in constructing a graph state. It
is realized by using a single hardware called the controlled-path (CP) gate
\cite{qlin1, qlin2,qlin3}. CP gate has been proposed as an element gate for
constructing different types of deterministic multi-qubit gates
\cite{qlin2,qlin3} and implementing quantum programs \cite{qlin4}. Its linear
optical version has been experimentally realized recently \cite{Brien}. The
use of the CP gates greatly simplified the generation of a special type of
graph states, the wheel-shape ones. On the other hand, with the growing number
of the involved qubits, the errors such as the missing of photons in
operations might occur. So far none of the previous proposals has considered
the detection of such error. The possible detection of operation error is
another feature in our study.

The rest of the paper is organized as follows. In Sec. II, in addition to a
brief introduction of CP gate, we will describe how to apply this gate for
implementing the basic operation for constructing graph states. In the
following two sections, we discuss how to generate graph states and detect
operation errors by the use of CP gates. Finally, in the last section, we
conclude the paper with a brief discussion.

\section{Performing cascade CZ operation with CP gate}

The controlled-Z (CZ) operation is generally considered for creating the
entanglement bonds between the qubits in a graph state \cite{Nielsen1, Browne,
Gilbert, dc-c1, experiments1, experiments2, k, pan1, y, pan2}. The independent
single photons can be entangled one after another by the repeated CZ
operations. These CZ operations can be, however, bundled together to the
so-called cascade CZ (CCZ) operation for speeding up the generation of a graph
state \cite{qlin5}. Here we illustrate an alternative way to realize such CCZ
operation with CP gates. The advantage of using CP gates will be shown in the
next section.

First of all, we present a brief introduction of the CP gate as our basic
hardware. The non-deterministic version of CP gate was first introduced in
\cite{qlin1}, and later it was developed in \cite{qlin2,qlin3} to the
deterministic one. A CP gate controls the spatial modes of a target photon by
the polarizations of the control photon. In generating graph states our target
photons initially carry two spatial modes, so we will apply the modified
version in Fig. 1 from its original design in \cite{qlin2,qlin3}.

Our modified CP gate implements the transformation
\begin{align}
\left\vert \Psi\right\rangle  &  =\left\vert H\right\rangle _{C}\left(
\frac{\left\vert \phi_{1}\right\rangle _{1}+\left\vert \phi_{1}\right\rangle
_{2}}{\sqrt{2}}\right)  +\left\vert V\right\rangle _{C}\left(  \frac
{\left\vert \phi_{2}\right\rangle _{1}+\left\vert \phi_{2}\right\rangle _{2}%
}{\sqrt{2}}\right) \nonumber\\
&  \rightarrow\left\vert H\right\rangle _{C}\left\vert \phi_{1}\right\rangle
_{1}+\left\vert V\right\rangle _{C}\left\vert \phi_{2}\right\rangle
_{2}=\left\vert \Phi\right\rangle ,
\end{align}
where $\left\vert \phi_{i}\right\rangle =\alpha_{i}\left\vert H\right\rangle
+\beta_{i}\left\vert V\right\rangle $, with $\underset{i=1}{\overset{2}{\sum}%
}( \left\vert \alpha_{i}\right\vert ^{2}+\left\vert \beta_{i}\right\vert ^{2})
=1$, are on two spatial paths denoted by $1, 2$. The effect of a CP gate
operation relates the spatial modes of the target photon to the polarization
modes of the control photon.

\begin{figure}[ptb]
\includegraphics[width=8.7cm]{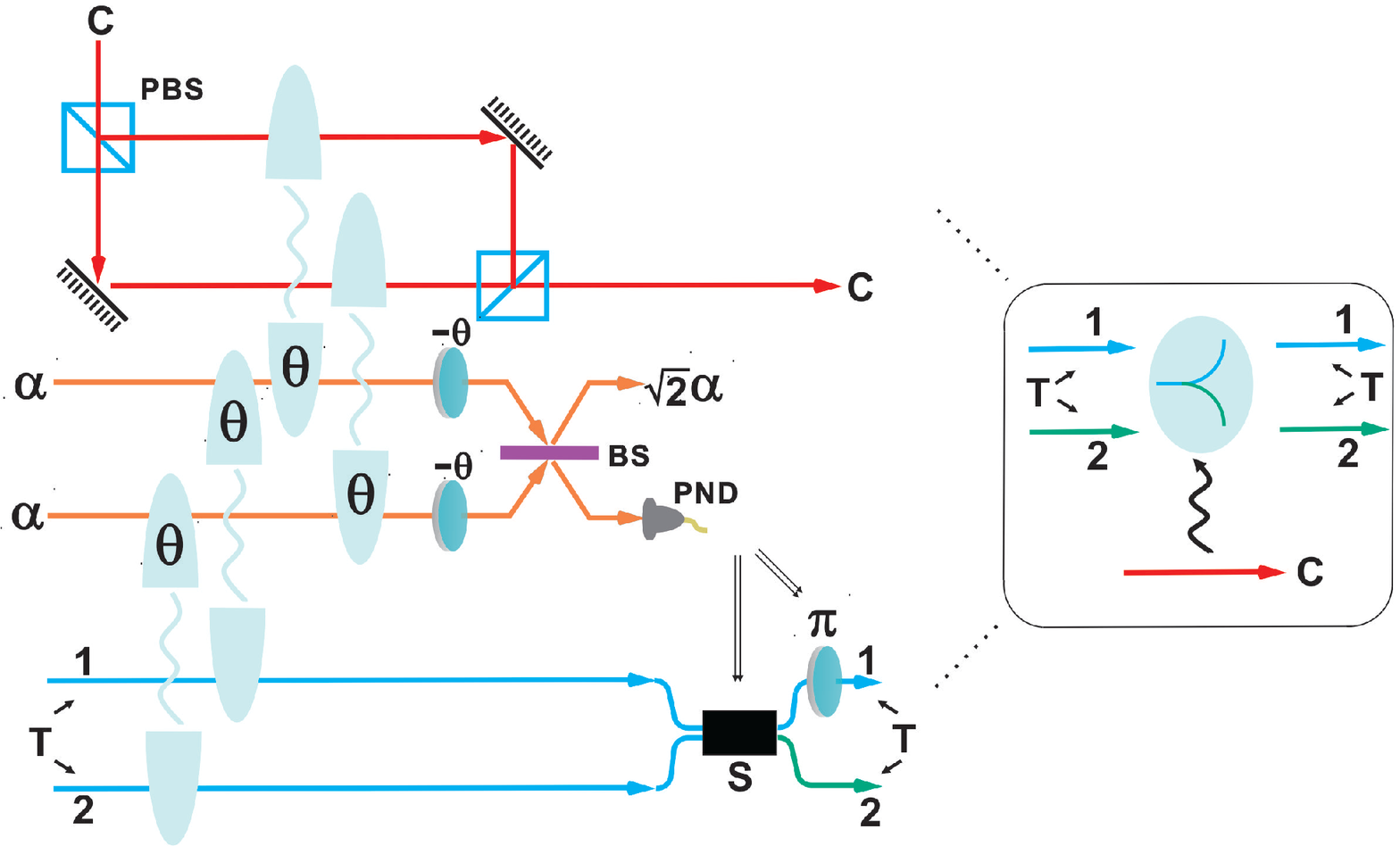}\caption{{ Setup of controlled-path
(CP) gate. Initially, the target photon of the input state carries two spatial
modes. The control photon and the target photon interact with each other
indirectly with their cross phase modulation processes with two ancilla
coherent states giving the conditional phase shifts $\theta$. After that, two
phase shifts and one beam splitter operation, associated with the photon
number detection and classical feed-forward control and a phase shift $\pi$,
complete the CP gate operation.}}%
\end{figure}

The core element to realize the above transformation is the XPM (cross phase
modulation) between the modes of the single photons and two coherent states;
see Fig. 1. Here the XPM based on weak cross-Kerr nonlinearity implements the
transformation $|1\rangle|\alpha\rangle\rightarrow|1\rangle|\alpha e^{i\theta
}\rangle$, with $\theta\ll1$, between a coherent state $|\alpha\rangle$ (or
called qubus beam) and a single-photon mode $|1\rangle$. The XPM configuration
in Fig. 1 transforms the input state to the following
\begin{align}
&  \frac{1}{\sqrt{2}}\left(  \left\vert H\right\rangle _{C}\left\vert \phi
_{1}\right\rangle _{1}+\left\vert V\right\rangle _{C}\left\vert \phi
_{2}\right\rangle _{2}\right)  \left\vert \alpha e^{i\theta}\right\rangle
\left\vert \alpha e^{i\theta}\right\rangle \nonumber\\
&  +\frac{1}{\sqrt{2}}\left\vert H\right\rangle _{C}\left\vert \phi
_{1}\right\rangle _{2}\left\vert \alpha\right\rangle \left\vert \alpha
e^{i2\theta}\right\rangle \nonumber\\
&  +\frac{1}{\sqrt{2}}\left\vert V\right\rangle _{C}\left\vert \phi
_{2}\right\rangle _{1}\left\vert \alpha e^{i2\theta}\right\rangle \left\vert
\alpha\right\rangle .
\end{align}
Then, applying the phase shifter of $-\theta$ to two coherent states and
implementing the transformation $\left\vert \alpha_{1}\right\rangle \left\vert
\alpha_{2}\right\rangle \rightarrow\left\vert \frac{\alpha_{1}-\alpha_{2}%
}{\sqrt{2}}\right\rangle \left\vert \frac{\alpha_{1}+\alpha_{2}}{\sqrt{2}%
}\right\rangle $ for them with a 50:50 beam splitter (BS), one will obtain the
state%
\begin{align}
&  \frac{1}{\sqrt{2}}\left(  \left\vert H\right\rangle _{C}\left\vert \phi
_{1}\right\rangle _{1}+\left\vert V\right\rangle _{C}\left\vert \phi
_{2}\right\rangle _{2}\right)  \left\vert 0\right\rangle \left\vert \sqrt
{2}\alpha\right\rangle \nonumber\\
&  +\frac{1}{\sqrt{2}}\left\vert H\right\rangle _{C}\left\vert \phi
_{1}\right\rangle _{2}\left\vert -\beta\right\rangle \left\vert \sqrt{2}%
\alpha\cos\theta\right\rangle \nonumber\\
&  +\frac{1}{\sqrt{2}}\left\vert V\right\rangle _{C}\left\vert \phi
_{2}\right\rangle _{1}\left\vert \beta\right\rangle \left\vert \sqrt{2}%
\alpha\cos\theta\right\rangle ,
\end{align}
where $\left\vert \beta\right\rangle =\left\vert i\sqrt{2}\alpha\sin
\theta\right\rangle $. By the projection $\left\vert n\right\rangle
\left\langle n\right\vert $ on the first qubus beam, one will obtain either
the target state
\begin{equation}
\left\vert H\right\rangle _{C}\left\vert \phi_{1}\right\rangle _{1}+\left\vert
V\right\rangle _{C}\left\vert \phi_{2}\right\rangle _{2}=\left\vert
\Phi\right\rangle \label{eq}%
\end{equation}
or the state
\begin{equation}
e^{-in\left(  \pi/2\right)  }\left\vert H\right\rangle _{C}\left\vert \phi
_{1}\right\rangle _{2}+e^{in\left(  \pi/2\right)  }\left\vert V\right\rangle
_{C}\left\vert \phi_{2}\right\rangle _{1},
\end{equation}
which can be transformed to the target state $\left\vert \Phi\right\rangle $
by the post operations in Fig. 1 according to the reading of $n$.

\begin{figure}[ptb]
\includegraphics[width=7.7cm]{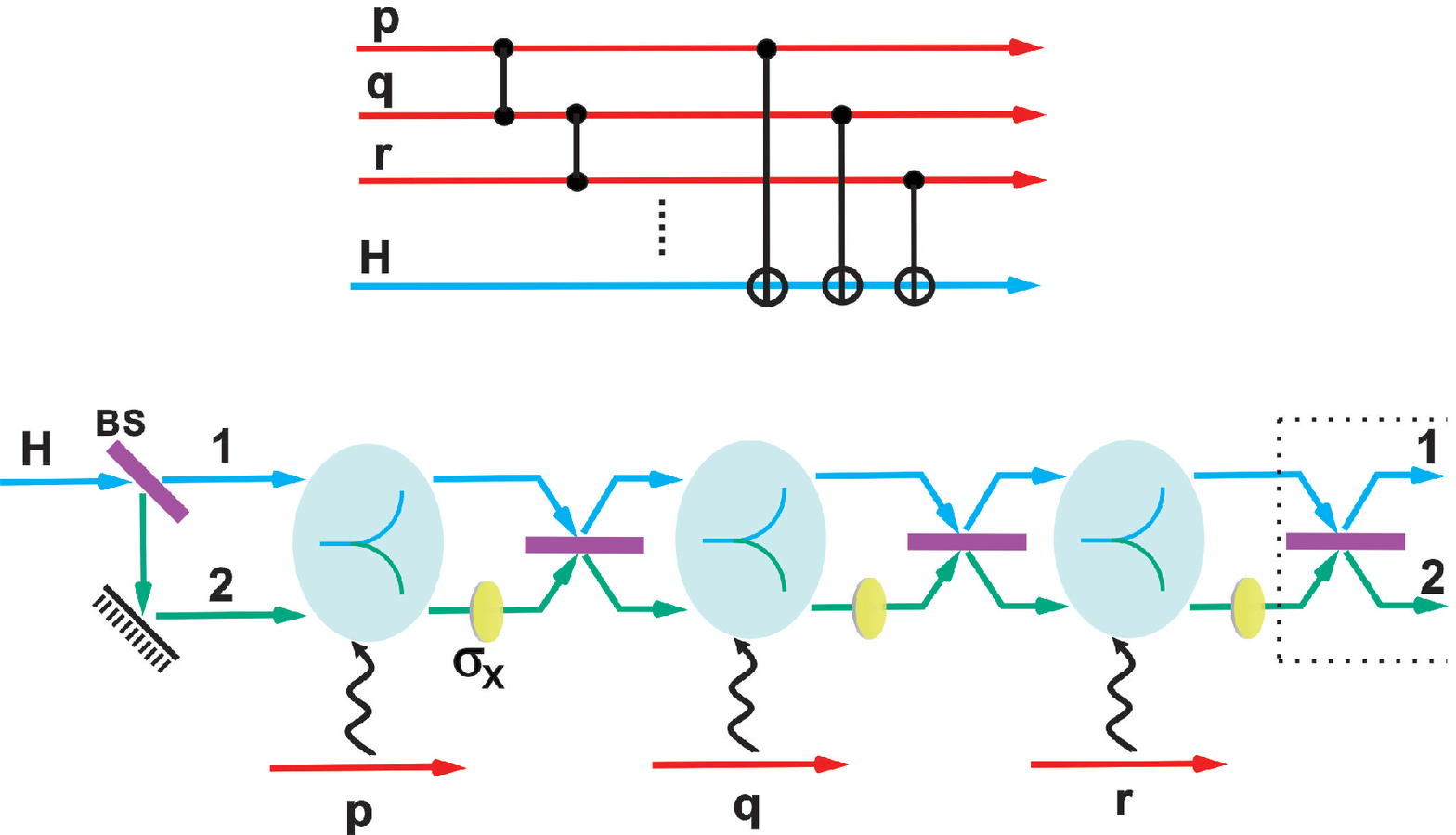}\caption{ { Realization of CCZ
operation with CP gate. The CCZ operations between the single photons $p,q,r$,
as well as the CNOT operations between the single photons $p,q,r$ and the
ancilla single photon in the state $\left\vert H\right\rangle $, can be
realized with three CP gates. One more beam splitter in the dashed line is
used to remove the unnecessary path information.}}%
\end{figure}

Now we apply CP gates to implement the CCZ operation in Fig.2. There are CZ
operations between single photon p and q, q and r, respectively. The direct
way to realize this operation should have two CZ operations one by one on the
three photons. The previous design of the circuit to implement a CZ operation
requires two entanglers together with one ancilla single photon \cite{Pittman,
Nemoto}. The so-called entangler transforms two single photons in the product
of the states $a\left\vert H\right\rangle +b\left\vert V\right\rangle $ and
$(\left\vert H\right\rangle +\left\vert V\right\rangle )/\sqrt{2}$ to the
entangled one in the state $a\left\vert HH\right\rangle +b\left\vert
VV\right\rangle $.\ However, it is not so efficient for doing so in realizing
a CCZ operation. In what follows we will show that a CCZ operation can be
implemented by CP gates. As it is shown in the above description for the CP
gate in Fig. 1, its complexity is equivalent to only one such entangler.

Without loss of generality, we use a three-photon CCZ operation as the example
(see Fig.2). The general form of the initial quantum state is
\begin{align}
&  \left\vert \psi_{1}\right\rangle \left\vert HHH\right\rangle +\left\vert
\psi_{2}\right\rangle \left\vert HHV\right\rangle +\left\vert \psi
_{3}\right\rangle \left\vert HVH\right\rangle \nonumber\\
&  +\left\vert \psi_{4}\right\rangle \left\vert HVV\right\rangle +\left\vert
\psi_{5}\right\rangle \left\vert VHH\right\rangle +\left\vert \psi
_{6}\right\rangle \left\vert VHV\right\rangle \nonumber\\
&  +\left\vert \psi_{7}\right\rangle \left\vert VVH\right\rangle +\left\vert
\psi_{8}\right\rangle \left\vert VVV\right\rangle ,
\end{align}
where $\left\vert \psi_{i}\right\rangle $ $\left(  i=1,\cdots,8\right)  $
denote the states of the other photons except of the concerned photons $p,q,r$.

In the first step, one introduces a single photon $\left\vert H\right\rangle $
as the ancilla and lets it be transmitted through a 50:50 BS to the state
$\left(  \left\vert H\right\rangle _{1}+\left\vert H\right\rangle _{2}\right)
/\sqrt{2}$. Applying the first CP gate (the photon $p$ controls the ancilla
photon), one will obtain the following state
\begin{align}
&  \left\vert \psi_{1}\right\rangle \left\vert HHH\right\rangle \left\vert
H\right\rangle _{1}+\left\vert \psi_{2}\right\rangle \left\vert
HHV\right\rangle \left\vert H\right\rangle _{1}+\left\vert \psi_{3}%
\right\rangle \left\vert HVH\right\rangle \left\vert H\right\rangle
_{1}\nonumber\\
&  +\left\vert \psi_{4}\right\rangle \left\vert HVV\right\rangle \left\vert
H\right\rangle _{1}+\left\vert \psi_{5}\right\rangle \left\vert
VHH\right\rangle \left\vert H\right\rangle _{2}+\left\vert \psi_{6}%
\right\rangle \left\vert VHV\right\rangle \left\vert H\right\rangle
_{2}\nonumber\\
&  +\left\vert \psi_{7}\right\rangle \left\vert VVH\right\rangle \left\vert
H\right\rangle _{2}+\left\vert \psi_{8}\right\rangle \left\vert
VVV\right\rangle \left\vert H\right\rangle _{2}.
\end{align}
Then a bit flip operation on the path mode 2 yields%
\begin{align}
&  \left\vert \psi_{1}\right\rangle \left\vert HHH\right\rangle \left\vert
H\right\rangle _{1}+\left\vert \psi_{2}\right\rangle \left\vert
HHV\right\rangle \left\vert H\right\rangle _{1}+\left\vert \psi_{3}%
\right\rangle \left\vert HVH\right\rangle \left\vert H\right\rangle
_{1}\nonumber\\
&  +\left\vert \psi_{4}\right\rangle \left\vert HVV\right\rangle \left\vert
H\right\rangle _{1}+\left\vert \psi_{5}\right\rangle \left\vert
VHH\right\rangle \left\vert V\right\rangle _{2}+\left\vert \psi_{6}%
\right\rangle \left\vert VHV\right\rangle \left\vert V\right\rangle
_{2}\nonumber\\
&  +\left\vert \psi_{7}\right\rangle \left\vert VVH\right\rangle \left\vert
V\right\rangle _{2}+\left\vert \psi_{8}\right\rangle \left\vert
VVV\right\rangle \left\vert V\right\rangle _{2}.
\end{align}

In the second step, the ancilla photon is transmitted to the second BS to
perform the operation
\begin{align}
\left\vert H\right\rangle _{1}  &  \rightarrow\left(  \left\vert
H\right\rangle _{1}+\left\vert H\right\rangle _{2}\right)  /\sqrt
{2}\nonumber\\
\left\vert V\right\rangle _{2}  &  \rightarrow\left(  \left\vert
V\right\rangle _{1}-\left\vert V\right\rangle _{2}\right)  /\sqrt{2}.
\end{align}
After that, the second CP gate operation is implemented (photon $q$ controls
the ancilla photon) to achieve the state%
\begin{align}
&  \left\vert \psi_{1}\right\rangle \left\vert HHH\right\rangle \left\vert
H\right\rangle _{1}+\left\vert \psi_{2}\right\rangle \left\vert
HHV\right\rangle \left\vert H\right\rangle _{1}+\left\vert \psi_{3}%
\right\rangle \left\vert HVH\right\rangle \left\vert H\right\rangle
_{2}\nonumber\\
&  +\left\vert \psi_{4}\right\rangle \left\vert HVV\right\rangle \left\vert
H\right\rangle _{2}+\left\vert \psi_{5}\right\rangle \left\vert
VHH\right\rangle \left\vert V\right\rangle _{1}+\left\vert \psi_{6}%
\right\rangle \left\vert VHV\right\rangle \left\vert V\right\rangle
_{1}\nonumber\\
&  -\left\vert \psi_{7}\right\rangle \left\vert VVH\right\rangle \left\vert
V\right\rangle _{2}-\left\vert \psi_{8}\right\rangle \left\vert
VVV\right\rangle \left\vert V\right\rangle _{2}.
\end{align}
Then, one more bit flip on the path mode 2 yields%
\begin{align}
&  \left\vert \psi_{1}\right\rangle \left\vert HHH\right\rangle \left\vert
H\right\rangle _{1}+\left\vert \psi_{2}\right\rangle \left\vert
HHV\right\rangle \left\vert H\right\rangle _{1}+\left\vert \psi_{3}%
\right\rangle \left\vert HVH\right\rangle \left\vert V\right\rangle
_{2}\nonumber\\
&  +\left\vert \psi_{4}\right\rangle \left\vert HVV\right\rangle \left\vert
V\right\rangle _{2}+\left\vert \psi_{5}\right\rangle \left\vert
VHH\right\rangle \left\vert V\right\rangle _{1}+\left\vert \psi_{6}%
\right\rangle \left\vert VHV\right\rangle \left\vert V\right\rangle
_{1}\nonumber\\
&  -\left\vert \psi_{7}\right\rangle \left\vert VVH\right\rangle \left\vert
H\right\rangle _{2}-\left\vert \psi_{8}\right\rangle \left\vert
VVV\right\rangle \left\vert H\right\rangle _{2}.
\end{align}

In the third step, the following operation
\begin{align}
\left\vert H\right\rangle _{1}  &  \rightarrow\left(  \left\vert
H\right\rangle _{1}+\left\vert H\right\rangle _{2}\right)  /\sqrt
{2}\nonumber\\
\left\vert V\right\rangle _{1}  &  \rightarrow\left(  \left\vert
V\right\rangle _{1}+\left\vert V\right\rangle _{2}\right)  /\sqrt
{2}\nonumber\\
\left\vert H\right\rangle _{2}  &  \rightarrow\left(  \left\vert
H\right\rangle _{1}-\left\vert H\right\rangle _{2}\right)  /\sqrt
{2}\nonumber\\
\left\vert V\right\rangle _{2}  &  \rightarrow\left(  \left\vert
V\right\rangle _{1}-\left\vert V\right\rangle _{2}\right)  /\sqrt{2}.
\label{ev}%
\end{align}
will be performed by the third BS. After the third CP gate operation (photon
$r$ controls the ancilla photon), the state%
\begin{align}
&  \left\vert \psi_{1}\right\rangle \left\vert HHH\right\rangle \left\vert
H\right\rangle _{1}+\left\vert \psi_{2}\right\rangle \left\vert
HHV\right\rangle \left\vert H\right\rangle _{2}+\left\vert \psi_{3}%
\right\rangle \left\vert HVH\right\rangle \left\vert V\right\rangle
_{1}\nonumber\\
&  -\left\vert \psi_{4}\right\rangle \left\vert HVV\right\rangle \left\vert
V\right\rangle _{2}+\left\vert \psi_{5}\right\rangle \left\vert
VHH\right\rangle \left\vert V\right\rangle _{1}+\left\vert \psi_{6}%
\right\rangle \left\vert VHV\right\rangle \left\vert V\right\rangle
_{2}\nonumber\\
&  -\left\vert \psi_{7}\right\rangle \left\vert VVH\right\rangle \left\vert
H\right\rangle _{1}+\left\vert \psi_{8}\right\rangle \left\vert
VVV\right\rangle \left\vert H\right\rangle _{2}%
\end{align}
will be obtained. Then, we will obtain the following state
\begin{align}
&  \left\vert \psi_{1}\right\rangle \left\vert HHH\right\rangle \left\vert
H\right\rangle _{1}+\left\vert \psi_{2}\right\rangle \left\vert
HHV\right\rangle \left\vert V\right\rangle _{2}+\left\vert \psi_{3}%
\right\rangle \left\vert HVH\right\rangle \left\vert V\right\rangle
_{1}\nonumber\\
&  -\left\vert \psi_{4}\right\rangle \left\vert HVV\right\rangle \left\vert
H\right\rangle _{2}+\left\vert \psi_{5}\right\rangle \left\vert
VHH\right\rangle \left\vert V\right\rangle _{1}+\left\vert \psi_{6}%
\right\rangle \left\vert VHV\right\rangle \left\vert H\right\rangle
_{2}\nonumber\\
&  -\left\vert \psi_{7}\right\rangle \left\vert VVH\right\rangle \left\vert
H\right\rangle _{1}+\left\vert \psi_{8}\right\rangle \left\vert
VVV\right\rangle \left\vert V\right\rangle _{2} \label{fi}%
\end{align}
by a bit flip on the path mode 2. The operation in each of the three steps
leading to the above state are similar. In the above state the polarizations
of the ancilla photon in the above depend on the parity of the three photons
($p,q,r$). Such effect is the repeated CNOT operations performed by photon
$p,q,r$ on the ancilla photon, respectively. For example, one $V$ mode of the
three photons in the second term of the above equation specifies the $V$ mode
of the ancilla in the same term, and the $H$ mode of the ancilla follows two
$V$ modes of the other photons in the sixth term, etc. Meanwhile, a CCZ
operation has been performed between the photons $p,q$ and $q,r$,
respectively, to obtain our target state. It is straightforward to generalize
the operation to the operations for $n$ single photons. A linear graph state
of $n$ photon can be straightforwardly created by the above process.

After the whole operation the ancilla photon should be removed without
changing anything else. It is the interference between its two spatial modes
on one more BS (shown in the dashed line), together with a Hadamard operation
on each path mode as follows:
\begin{align}
\left\vert H\right\rangle _{1}  &  \rightarrow\frac{1}{\sqrt{2}}\left(
\left\vert H\right\rangle _{1}+\left\vert H\right\rangle _{2}\right)
\rightarrow\frac{1}{2}\left(  \left\vert H\right\rangle _{1}+\left\vert
V\right\rangle _{1}+\left\vert H\right\rangle _{2}+\left\vert V\right\rangle
_{2}\right) \nonumber\\
\left\vert V\right\rangle _{1}  &  \rightarrow\frac{1}{\sqrt{2}}\left(
\left\vert V\right\rangle _{1}+\left\vert V\right\rangle _{2}\right)
\rightarrow\frac{1}{2}\left(  \left\vert H\right\rangle _{1}-\left\vert
V\right\rangle _{1}+\left\vert H\right\rangle _{2}-\left\vert V\right\rangle
_{2}\right) \nonumber\\
\left\vert H\right\rangle _{2}  &  \rightarrow\frac{1}{\sqrt{2}}\left(
\left\vert H\right\rangle _{1}-\left\vert H\right\rangle _{2}\right)
\rightarrow\frac{1}{2}\left(  \left\vert H\right\rangle _{1}+\left\vert
V\right\rangle _{1}-\left\vert H\right\rangle _{2}-\left\vert V\right\rangle
_{2}\right) \nonumber\\
\left\vert V\right\rangle _{2}  &  \rightarrow\frac{1}{\sqrt{2}}\left(
\left\vert V\right\rangle _{1}-\left\vert V\right\rangle _{2}\right)
\rightarrow\frac{1}{2}\left(  \left\vert H\right\rangle _{1}-\left\vert
V\right\rangle _{1}-\left\vert H\right\rangle _{2}+\left\vert V\right\rangle
_{2}\right)  .
\end{align}
These transformations specify the post operations to obtain the target state:
if the detection of ancilla photon on path 1 is $\left\vert H\right\rangle $,
we will get the desired state directly; if the detection on path 1 is
$\left\vert V\right\rangle $, the desired state could be also achieved by
implementing three phase flip $\sigma_{z}$ operations on photons $p,q,r$,
respectively; if the detection on path 2 is $\left\vert H\right\rangle $, a
phase flip $\sigma_{z}$ operation implemented on photon $r$ will yield the
desired state; if the detection on path 2 is $\left\vert V\right\rangle $, we
should implement two phase flip $\sigma_{z}$ operations on photon $p,q$,
respectively. Meanwhile, if the ancilla photon should be preserved, the
unnecessary path information can be removed by a merging gate introduced in
\cite{qlin2, qlin3}. The merging gate performs the inverse transformation of a
CP gate operation like the following:
\begin{equation}
\left\vert H\right\rangle _{C}\left\vert \phi_{1}\right\rangle _{1}+\left\vert
V\right\rangle _{C}\left\vert \phi_{2}\right\rangle _{2}\rightarrow\left\vert
H\right\rangle _{C}\left\vert \phi_{1}\right\rangle +\left\vert V\right\rangle
_{C}\left\vert \phi_{2}\right\rangle ,
\end{equation}
i.e. it merges the different spatial modes of a target photon into the same
spatial mode without changing anything else. Therefore the subscripts $1$ and
$2$ in Eq. (\ref{fi}) can be removed after a merging gate operation.

\section{Generation of graph states}

In what follows we will illustrate how to generate different types of graph
states with the CCZ operation described above.

\subsection{Wheel-shape states}

As it has been describe in the above, after the CCZ operation in Fig. 2, the
single photons used for qubits and the ancilla photon are correlated due to
the effective CNOT operations between them, so the ancilla photon carries the
parity information of the other single photons. If we preserve the ancilla
single photon by a merging gate \cite{qlin2, qlin3} and perform a Hadamard
gate operation on it, the ancilla single photon will be in the state
$\left\vert +\right\rangle $ ($\left\vert -\right\rangle $) while the other
single photons are in the even (odd) parity state. In other words, there exist
the entanglement bonds between the ancilla photon and the other single
photons. This property enables one to generate the wheel-shape graph states in
a rather efficient way.

In Fig. 3 we compare the generation of the wheel-shape graph states by the
present CZZ operation based on CP gates and the CZZ operation adopted in
\cite{qlin5}. Because the above-mentioned correlation between the ancilla
single photon and the other single photons is destroyed in the approach of
\cite{qlin5}, some of the initial building blocks have to be the two-photon
entangled states in the form $\left(  \left\vert 0\right\rangle \left\vert
+\right\rangle +\left\vert 1\right\rangle \left\vert -\right\rangle \right)
/\sqrt{2}$. There the wheel-shaped cluster state could be generated only after
the ancilla photon has walked through the red paths in Fig. 3. One sees from
the figure that the ancilla photon has to move from the single photons on the
boundary to the central single photon again and again. Actually these steps
are unnecessary in our present approach. Here the single photon at the wheel
center works as the ancilla photon in the CCZ operation and it moves on the
indicated red path to entangle all other photons together. In the present
approach, therefore, all building blocks can be the independent single photons
only. Moreover, the number of the operation steps is reduced to make the
generation of such graph states simpler.

\begin{figure}[ptb]
\includegraphics[width=7.7cm]{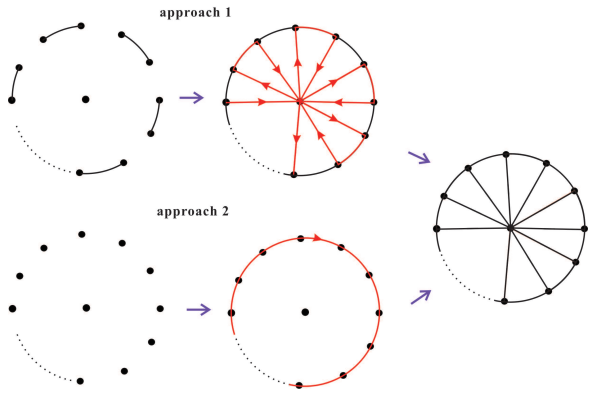}\caption{ { Comparisons of the
wheel-shape graph state generation in the approach of \cite{qlin5} and the
present approach using CP gates. The ancilla photon should walk from the
marginal single photons to the central single photon again and again in the
former approach, while it only moves through the marginal single photons in
the CP gate based method described in Sec. II.}}%
\end{figure}

\subsection{Two-dimensional and arbitrary graph state}

A universal 2D graph state is a 2D plane grid, which can be decomposed into a
series of broken lines and connected lines. In Fig. 4 we show one of such
decompositions, where six straight lines (green lines in step 1 and this type
linear graph states can be generated without ancilla \cite{qlin6}) and two
broken lines (red lines in step 2 and 4, and this type of graph states should
be created with the help of one ancilla single photon). Following the four
indicated steps here, one could generate a 2D cluster state (see the detailed
process in the next section). Such decomposition is not unique. The 2D graph
state will be constructed by connecting the indicated lines with CZZ
operation. For an arbitrary graph state in three dimension, we could also
decompose it into connected and unconnected lines in the most convenient way.

\section{Operation error detection}

\begin{figure}[ptb]
\includegraphics[width=7.7cm]{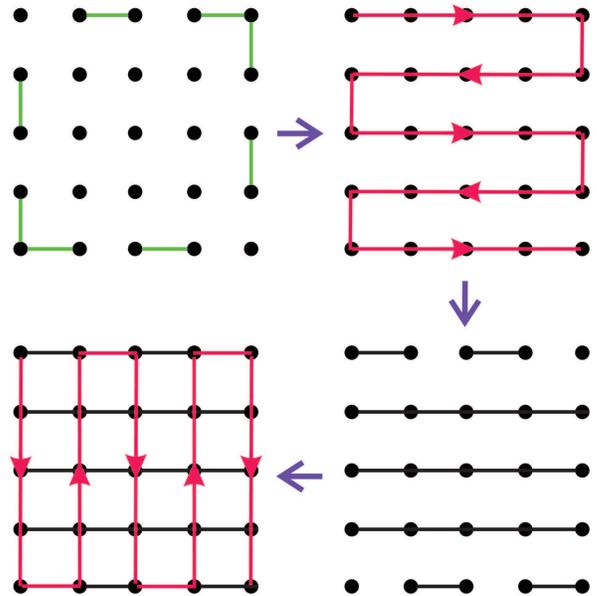}\caption{ { Generation of a
$5\times5$ 2D graph state with error detection. The starting blocks are linear
states and independent single photons. An ancilla single photon walks through
all the single photons along the red line in step 2 to reach the state in step
3. After that, one lets another ancilla photon walk through the same single
photons along anther red line in step 4 to generate $5\times5$ 2D graph state.
The comparison of the two ancilla single photons is used to detect whether all
entangling operations have been properly performed and it succeeds with a
probability of 1/2.}}%
\end{figure}

The CCZ operation proposed here also makes it possible to detect operation
errors. These error could arise from the missing of an operation on the input
photons. The larger the number of involved qubits, the more likely an
operation error could occur. Here we use the generation of a $5\times5$ graph
state to illustrate the point, and the detection method described below can be
used to other types of graph states as well. As it is shown in Fig.4, we
should first generate some linear states $\left(  \left\vert 0\right\rangle
\left\vert +\right\rangle +\left\vert 1\right\rangle \left\vert -\right\rangle
\right)  /\sqrt{2}$ and $\left(  \left\vert 0\right\rangle \left\vert
+\right\rangle \left\vert +\right\rangle +\left\vert 1\right\rangle \left\vert
-\right\rangle \left\vert -\right\rangle \right)  /\sqrt{2}$ while having some
independent single photons $\left\vert +\right\rangle $ as the initial
building blocks. These linear states can be generated without ancilla
\cite{qlin6}. Secondly, we use a single photon as the ancilla in the state
$\left\vert H\right\rangle $ (not show in the figure) and let it move through
all of the $n$ single photons (i.e., interact with the single photons as in
Fig.2) along the paths denoted by the red line in step 2 of Fig.4. The result
after the operation is the following state%
\begin{equation}
\left\vert even\right\rangle \left\vert H\right\rangle _{a_{1}}+\left\vert
odd\right\rangle \left\vert V\right\rangle _{a_{1}},
\end{equation}
where the $\left\vert even\right\rangle =\underset{l=0}{\overset{\left\lfloor
n/2\right\rfloor }{\sum}}f_{e}\left(  \theta\right)  P\left\vert
H^{\otimes(n-2l)}V^{\otimes2l}\right\rangle $ or $\left\vert odd\right\rangle
=\underset{l=0}{\overset{\left\lfloor n/2\right\rfloor }{\sum}}f_{o}\left(
\theta\right)  P\left\vert H^{\otimes(n-2l-1)}V^{\otimes\left(  2l+1\right)
}\right\rangle $. Here $\left\lfloor n/2\right\rfloor $ denotes the integer no more
than $n/2$, $P$ denotes a permutation operator, and $f_{e,o}\left(
\theta\right)$ is the corresponding prefactor for the even (odd) parity component of the other photons.
Thirdly, we introduce one more single photon as another ancilla in the state
$\left\vert H\right\rangle $ and also let it walk through all single photons
along the other path denote by the red line in step 4. Since the CZ operations
will not change the polarization $|H\rangle$ and $|V\rangle$ of the other
photons except for introduce extra phase of $\pi$ to the certain components,
we will achieve the state
\begin{equation}
\left\vert even^{^{\prime}}\right\rangle \left\vert H\right\rangle _{a_{1}%
}\left\vert H\right\rangle _{a_{2}}+\left\vert odd^{^{\prime}}\right\rangle
\left\vert V\right\rangle _{a_{1}}\left\vert V\right\rangle _{a_{2}%
},\label{p-state}%
\end{equation}
where the $\left\vert even^{^{\prime}}\right\rangle =\underset{l=0}%
{\overset{\left\lfloor n/2\right\rfloor }{\sum}}f_{e}^{^{\prime}}\left(
\theta\right)  P\left\vert H^{\otimes(n-2l)}V^{\otimes2l}\right\rangle $ and
$\left\vert odd^{^{\prime}}\right\rangle =\underset{l=0}{\overset{\left\lfloor
n/2\right\rfloor }{\sum}}f_{o}^{^{\prime}}\left(  \theta\right)  P\left\vert
H^{\otimes(n-2l-1)}V^{\otimes\left(  2l+1\right)  }\right\rangle $ denotes the
even or odd component, respectively.

Obviously the polarizations of the two ancilla single photons will be the
same, if all the operations are performed as desired. However, if one of the
operations miss the proper photons, the polarizations of the two ancilla
single photons will be different. Without loss of generality we suppose the
last CZ operation in step 4 has been missed. Then we will get the following
state%
\begin{align}
&  \left\vert even^{^{\prime\prime}}\right\rangle _{1}\left\vert
H\right\rangle _{l}\left\vert H\right\rangle _{a_{1}}\left\vert H\right\rangle
_{a_{2}}+\left\vert odd^{^{\prime\prime}}\right\rangle _{1}\left\vert
V\right\rangle _{l}\left\vert H\right\rangle _{a_{1}}\left\vert V\right\rangle
_{a_{2}}\nonumber\\
&  +\left\vert odd^{^{\prime\prime}}\right\rangle _{2}\left\vert
H\right\rangle _{l}\left\vert V\right\rangle _{a_{1}}\left\vert V\right\rangle
_{a_{2}}+\left\vert even^{^{\prime\prime}}\right\rangle _{2}\left\vert
V\right\rangle _{l}\left\vert V\right\rangle _{a_{1}}\left\vert H\right\rangle
_{a_{2}},\label{e-state}%
\end{align}
where subscript $\mathit{l}$ denotes the last single photon, and $\left\vert
even^{^{\prime\prime}}\right\rangle _{1}=\underset{l=0}{\overset{\left\lfloor
n/2\right\rfloor }{\sum}}f_{e}^{^{^{\prime\prime}}}\left(  \theta\right)
P\left\vert H^{\otimes(n-2l-1)}V^{\otimes2l}\right\rangle $, $\left\vert
odd^{^{\prime\prime}}\right\rangle _{1}=\underset{l=1}{\overset{\left\lfloor
n/2\right\rfloor }{\sum}}f_{e}^{^{^{\prime\prime}}}\left(  \theta\right)
P\left\vert H^{\otimes(n-2l)}V^{\otimes(2l-1)}\right\rangle $, $\left\vert
even^{^{\prime\prime}}\right\rangle _{2}=\underset{l=0}{\overset{\left\lfloor
n/2\right\rfloor }{\sum}}f_{o}^{^{^{\prime\prime}}}\left(  \theta\right)
P\left\vert H^{\otimes(n-2l-1)}V^{\otimes\left(  2l\right)  }\right\rangle $,
$\left\vert odd^{^{\prime\prime}}\right\rangle _{2}=\underset{l=0}%
{\overset{\left\lfloor n/2\right\rfloor }{\sum}}f_{o}^{^{^{\prime\prime}}%
}\left(  \theta\right)  P\left\vert H^{\otimes(n-2l-2)}V^{\otimes\left(
2l+1\right)  }\right\rangle$. The polarizations of the two ancilla single
photons in the second and four term of Eq. (\ref{e-state}) are different. Therefore, the
comparison of the polarizations of the second and four components will let us
know if all the single photons have been properly operated on or not.

This comparison will be done as follows. We first merge each ancilla single
photons into one spatial mode by a merging gate respectively and then let the
two ancilla photons interfere with each other via a PBS; see Fig.5(a). If the
polarizations are the same, one and only one single photon will appear on each
output modes of the PBS. Otherwise, two single photons will appear on one of
the output modes while no single photons will appear on the other output mode.
A coincident measurement of each output modes by the setups in the dashed line
of Fig. 5 is used to distinguish between the two cases. Here, before we detect
the ancilla single photon by single photon detectors (SPD), a Hadamard
operation, together with one more PBS placed on each output modes, is used to
preserve the proper quantum state superposition without projecting the
generated graph state onto its odd or even components.
The success probability for such error detection is determined by the parity of the polarization modes of the ancilla photons in Eq.(\ref{e-state}). The useful terms with the proper polarization of the two ancilla photons in (\ref{e-state}) occurs with the probability 0.5, fixing this value of detection success probability. 

\begin{figure}[ptb]
\includegraphics[width=7.0cm]{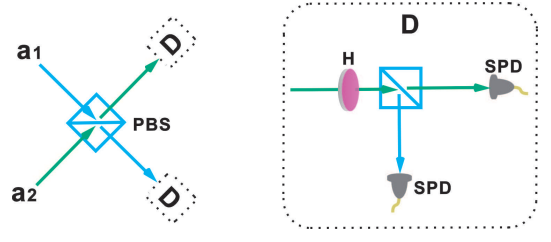}\caption{ Comparison of the
polarization of the ancilla photons. After the interference of two ancilla
photons via a PBS, the coincident measurement of the two output modes
indicates the comparison result. The detection setup in the dashed line
consists of a Hadamard operation (H), a PBS and two single photon detector
(SPD). The operations by these setups preserve the generated graph state.}%
\end{figure}

\section{Discussion and conclusion}

We have proposed a design based on CP gates to bundle the CZ operations for
generating graph states. The generation of a special type of graph states,
wheel-shape states, can be greatly simplified in the approach. Such setup
based on CP gates also allows to detect the possible operation error with a
success probability $1/2$.

As far as the ancilla photons are concerned, their main function is to assist
the generation of graph states. Like in our previous work \cite{qlin5}, an
ancilla photon acts as a spider, which weaves the independent single photons
into an arbitrary graph. A difference of the present approach using CP gates
is that the ancilla photon also carries the parity information of the other
single photons, so there is a possibility to detect the existence of operation
errors. The possible errors are manifested by the parity of a pair of ancilla
photons as in Eq. (\ref{e-state}) and the detection is achieved by the
coincident measure of the ancila photons' polarization modes, so one will be
limited to the success probability of 1/2 for such detection. Compared with
all other approaches, however, the use of CP gates allows one to find the
operation errors with a chance. The CP gates have been proposed as an element
for the circuit-based quantum computation \cite{qlin2,qlin3}. The same element
gate is used here for MBQC as well. The linear optical version of the gate has
been experimentally realized \cite{Brien}, making the principle-of-proof
demonstration of the present scheme feasible with the available technology.

In our discussion on the deterministic setups, we adopt the single-mode
treatment for the XPM between coherent states and single photons. This is a
valid approximation for the realistic multi-mode XPM under certain physical
conditions \cite{xpm}. We expect that the present proposal will be
experimentally realized with the proper XPM technology in the future.

\begin{acknowledgments}
The authors thank Ru-Bing Yang for helpful suggestions. Q. L. is supported by
National Natural Science Foundation of China (Grant No.11005040), Program for
New Century Excellent Talents in Fujian Province University (Grant No.
2012FJ-NCET-ZR04), the Fundamental Research Funds for Central Universities
(Grant No. JB-SJ1007), the Key Discipline Construction Project of Huaqiao
University and the State Scholarship Fund from China Scholarship Council. B.
H. acknowledges the support by AITF.
\end{acknowledgments}

\end{document}